\documentclass[11pt]{article}

\usepackage{graphics}
\usepackage{latexsym}
\usepackage{amsmath}
\usepackage{amssymb}
\usepackage[dvips]{epsfig}

\makeatletter

\def\singlespace{\def\baselinestretch{1}\@normalsize}

\newcommand{\doublespace}{\addtolength{\baselineskip}{0.10\baselineskip}}

\newcommand{\fig}[2]{
    \begin{figure}[thp]                        %
        \centerline{\psfig{figure=#1.eps,width=5.5in}}    %
        \small  
        \begin{singlespace}                               %
        \caption{#2 \label{#1}}                           %
        \end{singlespace}                                 %
        \end{figure}                          %
    }

\newcommand{\X}{\mbox{\boldmath$X$}}

\newcommand{\x}{\mbox{\boldmath$x$}}

\newcommand{\eps}{\varepsilon}

\def\t{\mbox{\boldmath $t$}}
\def\tt{\mbox{\boldmath $\scriptstyle t$}}

\def\x{\mbox{\X}}

\def\y{\mbox{\boldmath $y$}}
\def\XX{\mbox{\boldmath $\scriptstyle X$}}
\def\B{\mathcal{B}}
\def\BB{\mbox{\boldmath $\scriptstyle \beta$}}

\def\Th{\mbox{\boldmath $\theta$}}
\def\TTh{\mbox{\boldmath $\scriptstyle \theta$}}
\def\sgn{\mathrm{sgn}}
\def\ve{\varepsilon}
\def\E{\mbox{E}}

\def\P{\mbox{\bf P}}

\def\today{\ifcase\month\or
  January\or February\or March\or April\or May\or June\or
  July\or August\or September\or October\or November\or December\fi
  \space\number\day, \number\year}

\newcommand{\beq}{\begin{equation}}
\newcommand{\eeq}{\end{equation}}
\newcommand{\beqnn}{\begin{eqnarray*}}
\newcommand{\eeqnn}{\end{eqnarray*}}
\newcommand{\beqn}{\begin{eqnarray}}
\newcommand{\eeqn}{\end{eqnarray}}

\title{Non-convex penalized regression spline}

\author{Heng Peng \\ 
Department Mathematics, \\ The Hong Kong Baptist University}

\date{}

\begin{document}

\maketitle
\setlength{\parskip}{1\parskip}

\setcounter{page}{1}




\setlength{\baselineskip}{20pt plus2pt minus1pt} \doublespace









\begin{singlespace}
\begin{abstract}
Regression spline is a useful tool in nonparametric regression.
However, finding the optimal knot locations is a known difficult
problem. In this article, we introduce the Non-concave Penalized
Regression Spline. This proposal
method not only produces smoothing spline with optimal convergence
rate, but also can adaptively select optimal knots simultaneously.
It is insensitive to the number of origin knots.  The method's performance in a
simulation has been studied to compare the other methods. The
problem of how to choose smoothing parameters, i.e. penalty
parameters in the non-concave regression spline is addressed.
\end{abstract}
\end{singlespace}

\bigskip


\section{Introduction}

In recently, much attention has been attracted by the penalized
regression spline. This method takes advantage of the smoothing
spline and the regression spline to simplify computation and knots
selection procedures. The smoothing spline can be regarded as a
special case of penalized splines with a quadratic roughness
penalty function. Eiler and Marx (1996, 1998) used the quadratic
difference penalty function in penalized regression spline for
univariate and additive models. Mammen and van de Geer (1997)
proposed to regard the total variation of smoothing function as
the penalty function and studied some asymptotic properties for
this kind penalized spline. Ruppert and Carroll (1997, 2000)
considered some properties of the $L_1$-penalized regression
spline. They also considered variable penalty parameters used in
penalized regression splines. Wand (1999) and Aerts, Clasekens and
Wand (2002) studied some theory of the penalized regression spline
when penalties have quadratic forms.

As introduced by Chapter 1, the trade-off between smoothness and
flexibility of the regression spline is controlled by the number
and positions of knots.  Since Smith (1982) firstly used the
statistical variable selection technique to adaptively select
optimal knots in fitting splines, there are a lot of works along
this direction in adaptive regression spline lecture, such as
Friedman and Silverman's TURBO (1989), Friedman's MARS (1991),
Stone, {\sl et al.}'s POLYMARS (1997),  Luo and Wahba's HAS
(1997), Ruppert's MYOPIC (2001) etc. Usually, this kind algorithms
use stepwise procedures, such as forward or backward. These
approaches are very different from those used to establish the
traditional theory of regression spline. The gaps between the
theory and practice in the regression spline remain widely open.
Though most penalized regression spline procedures use convex
function as the penalty to produce shrinkage estimate to avoid the
choice of knots,  to get optimal smoothing for the function and
make the  procedures insensitive to knot number and locations,
these procedures cannot avoid involving many knots, a high
dimension parameter space, just like the smoothing spline. Hence
most penalized splines cannot take the full advantage of the
regression spline. In practice, the efficiency of those penalized
regression splines are still related to the optimal choice of the
number and locations of knots. Adaptive knots selection algorithms
still have to be studied for those penalized regression splines.
See for example, Ruppert and Carroll (1997), Mammen and van de
Geer (1997) and Ruppert (2002). There is little theoretical work
on penalized regression splines with various penalties.
Theoretical properties for penalized splines need further study.

In this paper, we propose a new approach, non-convex penalized
regression splines for spline fitting.  It is also easily extended
to multivariate function estimation problems. Unlike the
traditional penalized spline, our penalized spline considers a
non-convex function as the penalty and avoids stepwise procedure
to select knots. It can estimate smoothing function and adaptively
select knots simultaneously. It is insensitive to the number of
initial knots as long as it is large enough. This enables us to
study sampling properties of our penalized spline. In this
chapter, we show that our penalized regression spline has the
optimal convergence rate empirically by the simulation compared to
other nonparametric regression methods.

To overcome the inefficiency of traditional variable selection
procedures, most attractive properties of non-concave penalty were
introduced by Fan and Li (2001).  They demonstrated that penalized
least squares with non-concave penalty may produce threshold
estimate. In their article, they also showed how to select
significant variables and estimate their coefficients
simultaneously via non-concave penalized likelihood, which is
different to the traditional procedures.  In Chapter 2, we studied
some properties of nonconcave penalized likelihood estimate in
high dimensional situations. The results may extend to the
nonparametric regression. Stone, {\sl et al.}(1997) regarded the
regression spline model as an extended linear model with high
dimension. Under this idea, most variable selection methods can be
modified for knots selection, so does the non-convex penalized
least-squares approach. The basic idea of our approach just like
the idea used in non-concave penalized likelihood. By taking
advantages of the non-concave penalty function, especially
singularity at origin, we get threshold estimators by non-convex
penalized regression spline with appropriate spline basis, such as
the truncated power basis. By threshold criteria, some estimated
coefficient in a spline basis approximation are shrunk to zero.
This means that we can select knots automatically by the penalized
regression splines and avoid stepwise procedure. On the other
hand, the threshold estimate has the same effect as shrinkage
estimate to make trade-off between the flexibility and the
smoothness. This property of the threshold estimate has been used
in the wavelets analysis, for example, by Donoho and Johnstone
(1994), Antoniadis and Fan (2001). Therefore our approach can
select knots and estimate the smoothing spline simultaneously.
Eilers and Marx (1996) claimed without proof that the penalized
spline by quadratic difference penalty is not sensitive to the
number of knots if initial number of knots is large enough. The
number of knots used in smoothing spline is same as sample size
(see Green and Sliverman 1994). In these two approaches, the
trade-off between the smoothing and the flexibility is only
controlled by the regularized penalized parameters. From this
insight, our new proposed approach should be also expected to be
insensitive to the initial number of knots and the trade-off
between the smoothing and flexibility is mainly controlled by the
regularized parameters. We demonstrate this property by the
simulation in Section 3. The proper penalized parameters can be
selected by many methods, such as the fivefold cross-validation
proposed by Fan and Li (2001), the generalized cross-validation
method used in smoothing splines, see for example Green and
Silverman (1994), Wahba (1990) and BIC criterion (Schwarz 1978).
In this chapter, we consider two methods to select the penalized
parameter for the non-convex penalized regression spline.

Various algorithms mentioned in Chapter 2 to optimize a
high-dimension nonconcave likelihood function can be used for
non-convex regression spline. In this chapter, we apply the
modified Newton-Raphson algorithm proposed by Fan and Li (2001) to
our non-convex regression spline.

In Section 2, we introduce the non-convex penalized regression
spline. In Section 3,  numerical simulation results are
demonstrated. In Section 4, we give some discussions on how to our
approach to multivariate regression models.

\section{Non-convex penalized Regression Spline}

\subsection{Penalized regression splines}

Consider the nonparametric regression model as follows,
\begin{equation}
\label{s21} y_i=f(x_i)+\ve_i, \quad i=1,2,\ldots, n,
\end{equation}
where $x_i\in [0,1]$ are either deterministic or random design
points , the $\{\ve_i\}$ are independent random error with mean
zero and a constant variance $\sigma^2$, $f(x)$ is a smooth
regression function that want to be estimated.

\par To estimate function $f(x)$, we consider spline space $S(p,\t)$ with
knots
\begin{equation}
\label{s22}
 \t=\{ 0=t_0 < t_1 < \cdots < t_{k+1}=1\}.
\end{equation}
For $p \ge 2$, $S(p,\t)$ is defined as follows
$$
S(p,\t)=\{ s(x) \in C^{p-2}[0,1]: s(x) \mbox{ is a polynomial of
order $p$ on each subinterval $[t_i, t_{i+1}]$} \}
$$
When $p=1$, $S(p,\t)$ is the set of step functions with jumps at
the knots.

It is known that space $S(p,\t)$ is  a $k+p$ dimension linear
function space and truncated power function series
$$ \X_x = \{ 1, x, x^2,\ldots, x^{p-1}, (x-t_1)^{p-1}_+, \ldots
(x-t_k)^{p-1}_+ \} $$ forms a basis of $S(p,\t)$ (see de Boor
1978). Thus we may approximate $f(x)$ in model (\ref{s21}) by a
spline with form
\begin{equation}
\label{s23} f(x,\B)=\X_x \B = \beta_0+\beta_1 x+ \cdots +
\beta_{p-1}x^{p-1} + \sum\limits_{i=1}^k
\beta_{p+i-1}(x-t_i)_+^{p-1}.
\end{equation}
Hence, the nonparametric regression model (\ref{s21}) becomes a
classic high dimension linear regression model. Of course, we can
replace the truncated power basis by other bases of the spline
space, such as the B-spline basis, in (\ref{s23}). Here we like to
use the truncated power basis just because deleting a knot $t_j$
is equivalent to setting the coefficient $\beta_{p+j-1}$ to zero.
The variable selection procedure accords with knots selection.

The penalized regression spline is defined as the  minimizer
$\hat{f}(x,\B) \in S(p,\t)$ of the penalized least-squares problem
\begin{equation}
\label{s24}  \min\limits_{f(x, \BB) \in S(p,\tt)}
\sum\limits_{j=1}^n \{y_j-f(x_j,\B)\}^2 + n\sum\limits_{j=1}^k
p_{\lambda_n} (|w_{p+i-1}\beta_{p+i-1}|),
\end{equation}
where $p_{\lambda_n}(|\cdot|)$ is a penalty function and
$\lambda_n$ is the penalized parameter, $\{w_{j} \}$ are penalized
weights. The latter rescale or standardize the basis function in
(\ref{s23}) and transform them back to the original scale. Note
that we don't penalize the monomial terms $1,x, \ldots, x^{p-1}$
for sake of interpretability.

\subsection{Non-concave penalty functions}
Selection of  the penalty function  in (\ref{s24}) is  important
for knot selection. As discussed in Chapters 1 and 2, Fan and Li
(2001) studied the non-concave penalized likelihood for variable
selection. They showed that a good penalty function should result
in an estimator with three properties: (1) Unbiasedness, in which
there is no over-penalization of large parameters to avoid
unnecessary biases; (2) sparsity, as the resulting penalized
likelihood estimator should follow a thresholding rule so that
insignificant parameters can automatically be set to zero to
reduce model complexity; (3) continuity, to avoid instability in
model prediction, whereby the penalty function should be chosen
such that its corresponding penalized likelihood produces
continuous estimators of data. By the result of Fan and Li (2001),
the penalty functions satisfying sparsity and continuity must be
singular at the origin. The condition $p'_\lambda(|\beta|)=0$ for
large $|\beta|$ is a sufficient condition for unbiasedness for a
large true parameter.

The above three principles for penalty functions are also useful
in nonparametric regression, especially, in penalized regression
splines. Generally, the under smoothing of penalized regression
spline is caused by the excessive number of knots and this problem
is attenuated by convex penalties to produce shrinkage estimate of
coefficients of the basis functions such as the rough penalty used
in smoothing splines. Here the thresholding rule provides an
attractive alternative to reduce the problem of under smoothing.
We may reduce the number of knots adaptively by a thresholding
rule. On the other hand, the properties of unbiasedness and
continuity keep the smoothing and stability of the penalized
regression spline when we reduce the number of knots.

Fan and Li (2001) proposed the Smoothly Cipped Absolute Deviation
Penalty (SCAD). It is a non-concave penalty with singular at the
origin. To recall, it is defined as follows
\begin{equation}
\label{s25} p'_\lambda(\theta)=\lambda \Big \{ I(\theta \le
\lambda)+\frac{(a\lambda-\theta)_+}{(a-1)\lambda}I(\theta>\lambda)\Big
\}
\end{equation}
for some $a>2$ and $\theta>0$. By Fan (1997), the simple penalized
least-squares problems
\begin{equation}
\label{s26} (z-\theta)^2/2+p_\lambda(|\theta|)
\end{equation}
 with SCAD penalty yields the solution
\begin{equation}
\label{s27} \hat{\theta}=\left \{ \begin{array}{ll}
\sgn(z)(|z|-\lambda)_+, & \mbox{when} |z|\le 2\lambda, \\
\{(a-1)z-\sgn(z)a\lambda\}/(a-2), & \mbox{when} 2\lambda< |z| \le
a\lambda, \\
z, & \mbox{when} |z|>a\lambda
\end{array} \right.
\end{equation}
Hence by (\ref{s26}), the estimator that SCAD penalty results in
has the three properties discussed above. The discussion of other
penalty functions can be referred to Chapter 2 or Fan and Li
(2001) and Antoniadis and Fan (2001). Here we just use SCAD
penalty to show the basic idea of non-convex penalized regression
spline.

\subsection{An iterative algorithm}
We may directly apply SCAD penalty in the right side of
(\ref{s24}) to get a non-convex penalized regression spline.
However, it poses challenges to minimize (\ref{s24}), which is a
high-dimensional problem.  Here we follow a simple iterative
algorithm proposed by Fan and Li (2001).

Suppose we have an initial value $\B_0$ that is close to the
minimizer of the right side of (\ref{s24}). The SCAD penalty
function is singular at the origin, and it does not have
continuous first order derivatives. Thus the first step of the
algorithm is to check if the initial value of $\beta_{j0}, j=p,
\ldots, p+k-1$ equal to zero. If $\beta_{j0}$ is very close to
$0$, then set $\hat{\beta}_j=0$. Otherwise we consider the
following quadratic approximation
$$
\{p_{\lambda_n}(|\beta_j|)\}'=
p_{\lambda_n}'(|\beta_j|)\sgn(\beta_j) \thickapprox
\{p_{\lambda_n}'(|\beta_{j0}|)/|\beta_{j0}|\}\beta_j
$$
when $\beta_j \ne 0$. In other words,
$$
p_{\lambda_n}(|\beta_j|)\thickapprox p_{\lambda_n}(|\beta_{j0}|)
+\frac12
\{p_{\lambda_n}'(|\beta_{j0}|)/|\beta_{j0}|\}(\beta_j^2-\beta_{j0}^2),
\quad \mbox{for} \quad \beta_j \thickapprox \beta_{j0}.
$$
Let $\beta_{j_10}, \ldots,\beta_{j_d0}$ be the nonzero components
of $\B_0$,  In this step we also define,
$$
\Sigma_{\lambda_n}(\B_0)= \mathrm{diag} \{
p_{\lambda_n}'(|w_{j_1}\beta_{j_10}|)/|w_{j_1}\beta_{j_10}|,\ldots,
p_{\lambda_n}'(|w_{j_d}\beta_{j_d0}|)/|w_{j_d}\beta_{j_d0}| \}
$$
$$
\X_{x_i}(\B_0)=\{1,x_i,\ldots, x_i^{p-1},
(x_i-t_{j_1})^{p-1}_+,\ldots,(x_i-t_{j_d})^{p-1}_+ \}
$$
and
$$
\X_n=\{\X_{x_1}^T, \ldots, \X_{x_n}^T\}^T \quad \mbox{and} \quad
\X_n(\B_0)=\{\X_{x_1}^T(\B_0), \ldots, \X_{x_n}^T(\B_0)\}^T
$$

In the second step, we  compute the ridge regression
$$
\B_1=\{\X_n(\B_0)^T\X_n(\B_0)+n\Sigma_{\lambda_n}(\B_0)\}^{-1}\X_n^T(\B_0)\y.
$$

The third step of the algorithm is updating $\B_0$ by $\B_1$ and
repeating the first and second steps until the the iterative
solution is numerically stable.

\subsection{Issues on practical implementation}
Fan and Li (2001) demonstrated the convergence of this algorithm
by simulation. Their test also indicates this algorithm converges
quickly. A drawback of this algorithm is that once a coefficient
is shrunken to zero, it will stay at zero. However, this drawback
also significantly reduces the computational burden.

We have claimed that the non-convex penalized regression spline is
insensitive to the number of the origin knots $k$ if it is large
enough. In the following simulation results, we can see that we
only require $k$ is large enough, generally no less than
$O(n^\frac{1}{2p+1})$ to get optimal convergence rate for the
non-convex penalized regression spline, where $n$ is the sample
size.  The knots series is defined as $t_i=x_{([ni/(k+1)])},
i=1,\ldots,k$, where $x_{(j)}$ is the $j$th order statistics of
$x_i$.

In practice, we do not know the smoothness conditions of the
function that we want estimate. To guard the efficiency of our
procedure to handle some spatial inhomogeneity setting, we follows
suggestion of the Friedman and Silverman (1989) that `` A smoother
should be resistant to a run of length $L$ of either positive or
negative error so long as its span in the region of the run is
large compared to $L$". Hence we would like to adopt the formula
proposed by Friedman and Silverman (1989) to select the value of
the minimum value of $k$,
\begin{equation}
\label{s28} k=[n/M(n,\alpha)]+1
\end{equation}
where $0.05 \le \alpha \le 0.1$ and  $n\ge 15$,
$M(n,\alpha)\approx L_{\max}(\alpha)/3$ (or $L_{\max}(\alpha)/2.5$
to be conservative) denotes a minimum span between two placed
knots, and $L_{\max}(\alpha)$ is the largest positive or negative
run to be expected in $n$ binomial trials with probability
$\alpha$. By numerical approximation,
\[
L_{\max}(\alpha)\approx -\log_2\{-(1/n)\ln(1-\alpha)\}.
\]

Our procedure can also be regarded as a backward algorithm for
selecting knots. In some sense, it only reduces the flexibility of
the model by deleting unwanted knots. Hence we hope that our
initial spline $\X_n\B_0$ has small bias and keep it under our
procedure. When the number of knots is large enough, it is obvious
that the least-squares estimate of regression spline is a good
choice. On the other hand, in Fan and Li (2001), they require that
the matrix $n^{-1} \X_n^T\X_n$ is not singular. In nonparametric
regression setting, $n^{-1}\X_n^T\X_n$ can be asymptotic singular,
specially when we use the truncated power basis as the basis of
the spline space, since the order of the matrix grows with $n$. In
this phase the initial estimate $\beta_{j0}$ obtained by the
regression spline may has a large variance. Thus we have to weigh
the initial estimate in penalty term by weight $w_j$ such that the
variance of $w_j\beta_{j0}$ is the order of $O(n^{-\frac12})$, the
same order as $\lambda_n$. Here, we take
$$
w_j=\Big [\big (\frac1n \X_n^T\X_n \big )^{-1}_{jj} \Big
]^{-\frac12}.
$$
In theory, we may able to inverse the above matrix , but in
practice, we replace the inverse operation by the generalized
inverse operation. In the classic linear model, if $n^{-1}
\X_n^T\X_n$ is singular, then the least squares estimate
$\hat{\B}=(\X_n^T\X_n)^-\X_n\y$ may not be a consistent estimate
of $\B$, but $\X_n\hat{\B}$ can still be a consistent estimable of
$\X_n\B$. Our simulation also show that though the truncated power
basis results in $n^{-1}\X_n^T\X_n$ that is asymptotically
singular, this has little influence on our numerical results.

\subsection{Selection of penalized parameters}
To implement our procedure, it is more important to estimate the
parameters $a$ and $\lambda_n$ for the SCAD than to decide the
value of $k$. $(a,\lambda_n)$ can be regarded as either smoothing
parameters or penalty parameters. We denote them by
$\Th=(a,\lambda_n)$. Here we discuss two methods of estimating
$\Th$: Predictor Risk Estimation Criterion (PREC) (often referred
to as the  $C_p$ Criterion) suggested by Eubank (1999), Modified
Generalized Cross-Validation (MGCV) proposed by Fan and Li (2001).

Let us first consider the modified generalized cross-validation.
In our iterative algorithm, we update the estimate $\B$ by
\begin{equation}
\label{s214}
\B_1(\Th)=\{\X_n(\B_0)^T\X_n(\B_0)+n\Sigma_{\lambda_n}(\B_0)\}^{-1}\X_n^T(\B_0)\y.
\end{equation}
Thus the fitted value of $\hat{f}(x_i)$ of $f(x_i),,i=1,\ldots,n $
is
$$
\X_n(\B_0)\{\X_n(\B_0)^T\X_n(\B_0)+n\Sigma_{\lambda_n}(\B_0)\}^{-1}\X_n^T(\B_0)\y,
$$
and the projection matrix can be defined as
$$
\P_{\XX_n}\{\hat{\B}(\Th)\}=\X_n(\hat{\B})\{\X_n(\hat{\B})^T
\X_n(\hat{\B})+n\Sigma_{\lambda_n}(\hat{\B})\}^{-1}\X_n^T(\hat{\B})
$$
Define the number of effective parameters in the non-convex
penalized regression spline as $e(\Th)
=\mathrm{tr}[\P_{\XX_n}\{\hat{\B}(\Th)\}]$. Hence, the modified
generalized cross-validation statistic is
\begin{equation}
\label{s215} \mathrm
{MGCV}(\Th)=\frac{1}{n}\frac{\|\y-X_n(\hat{\B})\hat{\B}(\Th)\|^2}
 {\{1-\gamma e(\Th)/n\}^2}
\end{equation}
where $\hat{\Th} =\mathrm{argmin}_{\TTh}\{\mathrm{MGCV}(\Th)\}$
and $\gamma$ is the inflated factor to be specified.

The predictor risk estimation criterion is defined as follows
\begin{equation}
\label{s216}
\hat{\P}\{\hat{\B}(\Th)\}=\frac{1}{n}\|\y-X_n(\hat{\B})\hat{\B}(\Th)\|^2+
\frac{2\gamma \sigma^2}{n} e(\Th),
\end{equation}
where $\hat{\Th}= \mathrm{argmin}_{\TTh}
\hat{\P}\{\hat{\B}(\Th)\}$, and $\gamma$ is also the inflated
factor.

The inflation factor used here is due to the fact that the a lot
of basis functions used in the model are selected adaptively (see
Luo and Wahba 1997 and Friedman 1991). When $\gamma=1$, the MGCV
is no difference to the GCV as suggested by Fan and Li (2001),
Breiman (1995), Tibshirani (1996), and Fu (1998). The predictor
risk estimation criterion also appears to Akaike's information
criterion, AIC (Akaike 1973). When $\gamma=\log (n)/2$, the
predictor risk estimation criterion is the Bayesian information
criterion, BIC, (Schwarz 1978).  In MGCV, following discussion of
Luo and Wahba (1997) and Friedman (1991), we suggest $\gamma $
should be in $ [1.2,3.5]$ to keep the stable of the non-convex
penalized regression spline. For the predictor risk estimation
criterion, our criterion is similar to the one proposed by Rao and
Wu (1989) used for the model selection in a classic regression
problem. By the strong consistent results of Rao and Wu (1989) and
Bai, Rao and Wu (1999),  the value of $\gamma$ can be selected
from a large range. It is only required that $\gamma/\log\log n
\to \infty$ and $\gamma=o(n)$.  Thus we tend to agree that the
predictor risk estimation criterion is stable for a large range
value of $\gamma$. But conservatively, we will select the value of
$\gamma$ from $[2,5]$ or the form of $\log (n)/2$. A lot model
selection criterions mentioned in Rao and Wu (1989) and Bai, Rao
and Wu (1999) can be also used here to select the smoothing
parameter $\Th$.

In fact, as shown by Fan and Li (2001), 3.7 is a good choice for
the parameter $a$ used in the SCAD. Hence, we mainly use the MGCV
and PREC to select the penalized parameter $\lambda_n$ for the
SCAD penalized regression spline. In the following simulation or
discussion, we always set the value of $a$ as 3.7.

\section{Simulation study}
In this section, we use the following 4 examples. The first two
come from Fan and Gijbels (1995), the last two come from Donohon
and Johnstone (1994).
\begin{table}[h]   \caption{Specifications of Simulation Examples}
\tabcolsep 0.1cm \label{T1} \center
\begin{tabular}{cccc cccccccc}
\hline \hline
          & &      & &         &  & Sample  & &                 & & Number of  \\
  Example & & f(x) & & $\sigma$ & & size (n) & & $SD(f)/\sigma$ & & replicates
 \\ \hline
   1     & & $\sin(2(4x-2)+2\exp(-16x^2)$ & & .3 & & 256 & & 2.80 & &
   400 \\
   2     & & $(4x-2)+2\exp(-(16(x-0.5))^2)$ & & .4 & & 256 & & 3.16 &
   & 400 \\
   3  & & $2.2(4*\sin 4\pi t -\sgn(t-0.3)-\sgn (0.72-t))$ & & 1.0
   & & 2048 & & 6.54 & & 31 \\
   4 & &  $22\{ t(1-t)\}^{\frac12} \sin \{2\pi
   (1+0.05)/(t+0.05)\}$ & & 1.0 & & 2048 & & 6.36 & & 31 \\
   \hline
\end{tabular}
\end{table}

\subsection{ MSE Compared to other methods}
 In this section, we use simulation examples to examine the
 performance of the SCAD PRS  and compare it with HAS (Luo and Wahba
 1997), MARS (Friedman 1991), wavelet shrinkage (SUREShrink,
 Donoho and Johnstone 1994), smoothing splines SS procedures,
 Local PS and Globe PS (Ruppert and Carroll 2000).
 The simulation results of other methods are excerpted from
 Luo and Wahba (1997) and Ruppert and Carroll(2000).

 Luo and Wahba (1997) used the pseudostandard normal random number
 generator {\em rnor}, a Fortran subroutine from
 CMLIB. For SUREShrink, Luo and Wabha used the software wavethresh,
 developed by Nason and Silverman (1994) in S-PLUS. They chose the
 ``primary  resolution level" as 5 and the wavelets family  {\em ``
 DaubLeAsymm with filter number 8"}. The Fortran routines used to
 compute HAS estimates. The smoothing splines (SS) are computed by
 using the code GCVSPL in Fortran by Woltring's code, with the
 smoothing parameters chosen by GCV. The code {\em mars3.5} was
 used for MARS. Local PS and Globe PS were computed by using matlab
 code which were programmed by Ruppert and Carroll (2000).

 To compare HAS, we select 60 initial knots for our first two
 examples. To test our rule for selecting the initial number of
 knots, we follow formula (\ref{s28}) and take the value
\begin{equation}
\label{s31} \left [\frac{3n}{-\log_2\{(-1/n)\ln(1-0.1)
\}}\right]+1
\end{equation}
For the first two examples, the initial knots number is nearly 60
for $n=256$ and for the last two examples are 432 for $n=2048$.
The regularization parameter $\lambda_n$ is selected by MGCV with
the inflation factor 2.5.

In the second, the third and the fourth example, we relax some
requirement for the tuning parameters used in our algorithm to
improve the speed of our algorithm and to observe if our algorithm
is stable for these tuning parameters. The medians of MSEs are
presented in Table 3.2. The SCAD PRS fits with median performance
are shown in Figure 3.1. They accord quite well with the true
regression function.

\begin{table}[h] \caption{ Median of MSE$\times1000$
with different methods and the Interquartile range of
MSE$\times1000$ (in parentheses)  } \tabcolsep 0.05cm \label{T2}
\center
\begin{tabular}{cccc cccccccccccc}
\hline \hline
 Example & & SCAD PRS & & HAS & & SS & & SUREShrink & & MARS & &
 Local PS  & & Global PS
 \\ \hline
   1     & & 5.4(3.1) & &7(6)  & & 6 (3) &  &18(4)
   & &    7(4)  & & 5.3(3.5) & & 6.1 (2.9) \\
   2     & & 9.3(5.3) & &12(11)  & & 10 (5)  & & 42(12)  &
   & 12 (7) \\
   3 & &  51 (8)  & &  39 (13) & & 75(5) & & 62(7) & &
   150 (14) \\
   4 & & 196 (20.3) & & 68 (15) & & 205 (11) & & 149 (13) \\
   \hline
\end{tabular}
\end{table}

\fig{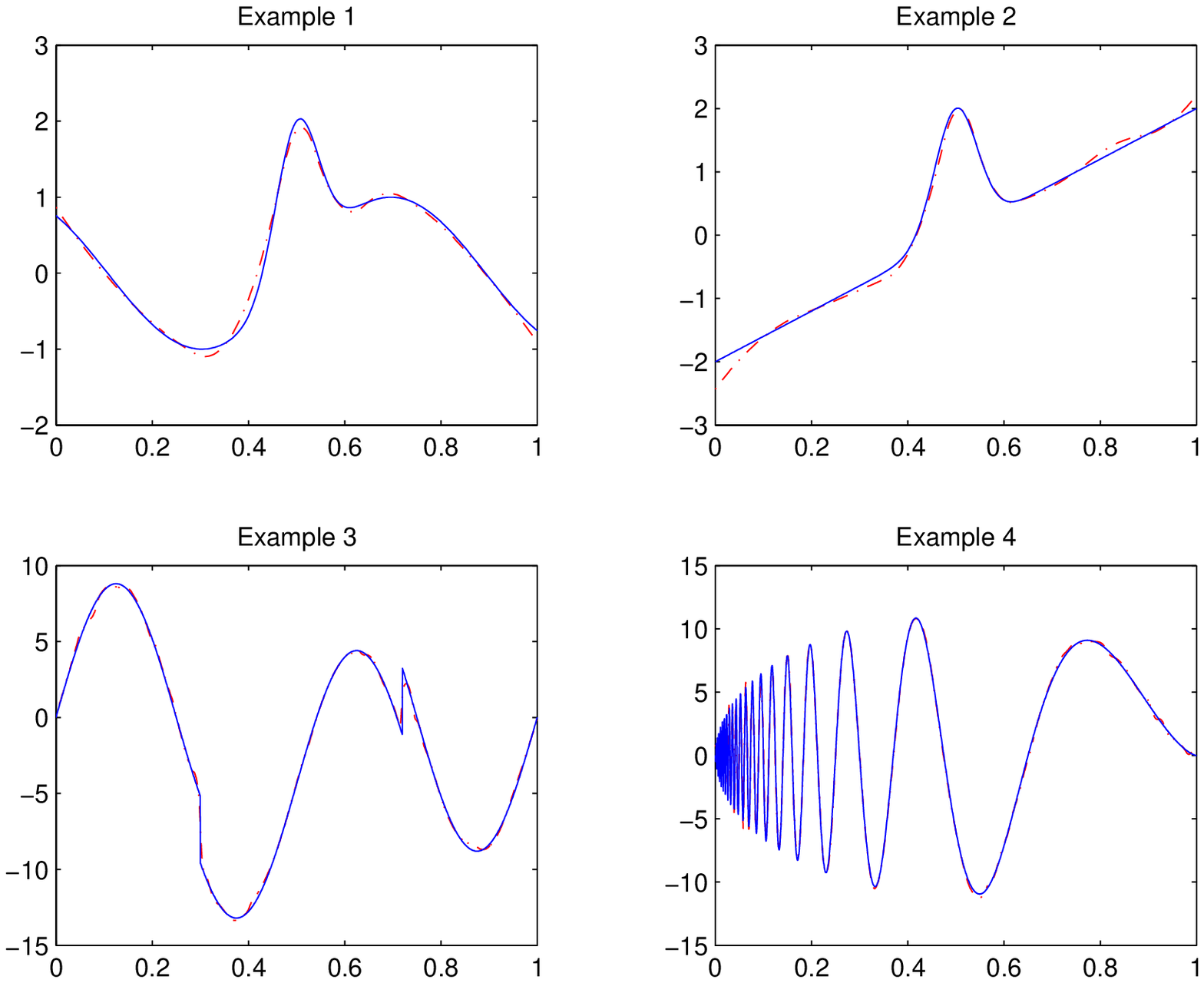}{SCAD PRS with Median of MSE for Examples
1-4 (solid line for true function, dot-dash for the estimate
function) }

In our simulation,  we use the quadratic spline as the results are
slightly better than those of cubic spline. This phenomenon was
also observed by Ruppert and Carroll (1999) when they study their
penalized regression spline with quadratic penalty.

From Table 3.2, our procedures is the best in first two examples,
and in the last two examples, our procedure slightly outperform
the smoothing spline, but not the HAS or SUREShrink. This is due
to the spatial inhomogeneity of the simulated function, for which
the HAS and SUREShrink are designed.

\subsection{Effect of initial number of knots for SCAD PRS}
 In this section, we study the influence of the initial number of
 knots used in SCAD penalized regression splines for the first two
 examples. We vary the initial number of knots from 30 to 150 with
 step size 30.  The median MSEs of 400 simulations are
 presented in Table 3.3. They show that the method
 is  insensitive to the number of initial knots though too many initial knots
 may cause slightly over fit.

\begin{table}[h]   \caption{ Median of MSE$\times 1000$ for different initial knots and
the Interquartile range of MSE$\times 1000$ (in parentheses)}
\tabcolsep 0.10cm \label{T3} \center
\begin{tabular}{cccc cccccccc}
\hline \hline {\small Initial Knots Number} & & 30 & & 60 & & 90 &
& 120  & & 150
 \\ \hline
   Example 1   & & 5.5(2.9) & & 5.4(3.1) & &  5.1(2.9) & & 5.2(2.7) & & 5.8(2.7) \\
   Example 2   & &  8.9(5.2)  & & 9.3(5.3) & &   9.4(5.1)& & 9.5(5.1)& & 9.8(5.2)  \\
   \hline
\end{tabular}
\end{table}

\subsection{Effect of knots selecting for SCAD PRS}
Depicted in Figures 3.2 and 3.3 are the locateions of the knots
finally automatically selected by non-convex penalized regression
spline. Figure 3.3 gives the frequency of every knot finally
selected by SCAD PRS in the 400 simulations for Examples 1 and 2.
Figures 3.3 gives the histograms for the number of knots that the
SCAD PRS finally selected in every simulation. It is obvious that
SCAD PRS used more knots than SCAD PRS of Example 1. This is
because we relax the requirement for the tuning parameters in our
algorithm. Hence, though these tuning parameters have little
influence on MSE, they affect the knots selection of SCAD PRS.

\fig{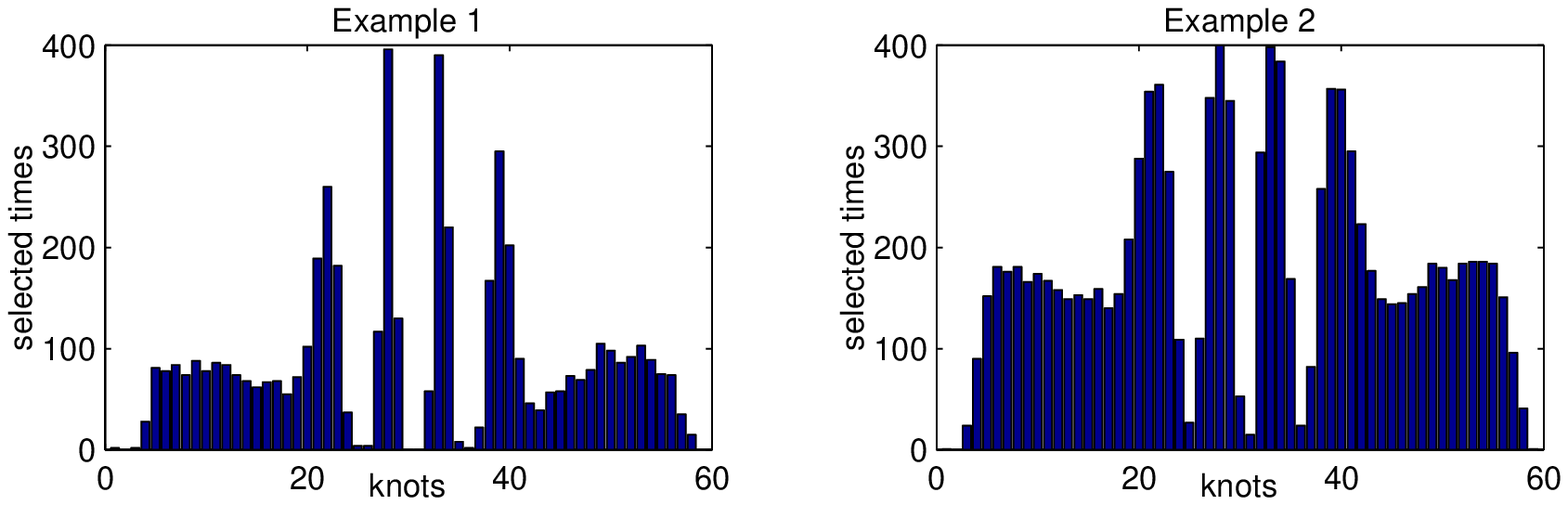}{Frequencies of the initial 60 knots that
are selected by SCAD PRS}

\fig{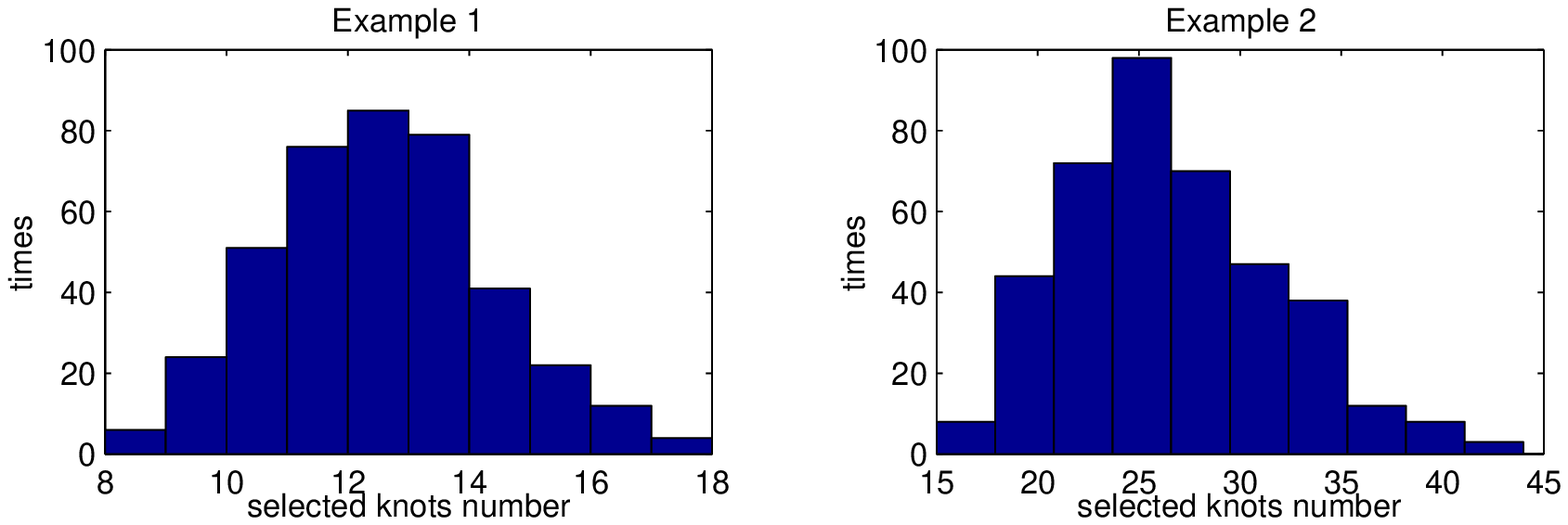}{The distributions of the knots selected by
the SCAD PRS with 60 initial knots}

\subsection{Parameters Selection Methods}
In this section, we mainly examine the performance of MGCV and
PREC with different number of initial knots. Specifically, we aim
at examining the impact of the inflation factor to the MGCV and
PREC on the estimated curves. For simplicity, we focus only on
Examples 1 and 2.

From Table 3.4-3.7, the two parameter selection methods, MGCV and
PREC are efficient in selecting the penalized parameter for the
non-convex penalized regression spline. However, MGCV method is
sensitive to the inflation factor. The best value of the inflation
factor for MGCV is in the interval $[2.0, 3.5]$.  The PREC method
is robust when the inflation factor is large. This result is
consistent with the results of Rao and Wu (1989) and Bai, Rao and
Wu (1999). The number of initial knots and the sample size may
also slightly affect the choice of the inflation factor. It needs
further study.

\begin{table}[h] \caption{ Median$\times 1000$ of MSE for different initial knots with MGCV Method and the Difference of
the First and  Third Quartiles of MSE (in Parentheses) for Example
1 } \tabcolsep 0.05cm  \center
\begin{tabular}{cccc cccccccccccccccc}
\hline \hline

  & & & \multicolumn{9}{c}{Initial Knots Number} &  \\ \cline{3-12}
   &Inflation factor $\gamma$ & &{30} & & {60} & &
  {90} & & {120}  & &
  {150}\\
\hline

& 1.0 &   &6.7(4.4)& &9.0(7.7)& & 11.0(13.0)&
          &14.4(18.6)& &23.4(25.9) \\

& 1.2 &   &6.2(3.8)& &6.7(4.5)& &6.3(5.5)&
          &6.1(4.9)& &6.6(5.5) \\

& 1.5 &   &5.8(3.4)& &5.5(3.2)& &5.3(3.0)&
          &5.4(3.0)& &5.9(3.0) \\

& 2.0 &   &5.5(3.0)& &5.3(3.0)& &5.2(2.8)&
          &5.2(2.8)& &5.8(2.8) \\

& 2.5 &   &5.5(2.9)& &5.4(3.1)& &5.1(2.9)&
          &5.2(2.7)& &5.8(2.7) \\

& 3.0 &   &5.6(2.9)& &5.6(32)& &5.2(2.9)&
          &5.2(2.8)& &52.4(14.0) \\

& 3.5 &   &5.7(3.0)& &5.8(3.3)& &5.2(2.9)&
          &5.2(2.8)& &53.8(8.3)\\

& 7.0 &   &67(42)& &72(42)& &32.6(6.7)&
          &43.3(7.8)& &53.8(8.3)\\

&$\ln(n)/2$&  &5.6(2.9)& &5.5(3.0)& &5.2(2.9)&
            &5.2(2.8)& &6.2(3.6) \\

&$\ln(n)$ &   &6.2(3.4)& &6.8(4.0)&
              &32.5(7.0)& &43.3(7.8)& &53.8(8.3) \\

&$\ln(k)/2$&   &5.7(31)& &5.3(3.1)& &5.2(2.9)&
               &5.2(27)& &5.8(2.7) \\

&$\ln(k)$ &    &5.7(3.0)& &6.0(3.6)& &5.2(2.9)&
               &43.3(7.8)& &53.8(8.3)
\\
\hline
\end{tabular}
\end{table}

\begin{table}[h]  \caption{ Median$\times 1000$ of MSE for different initial knots with PREC Method and the Difference of
the First and  Third Quartiles of MSE (in Parentheses) for Example
1 } \tabcolsep 0.05cm  \center
\begin{tabular}{cccc cccccccccccccccc}
\hline \hline

  & & & \multicolumn{9}{c}{Initial Knots Number} &  \\ \cline{3-12}
   &Inflation factor $\gamma$ & &{30} & & {60} & &
  {90} & & {120}  & &
  {150}\\
\hline
 &1.0&  &6.8(4.4)& &10.4(8.2)& &15.7(14.1)&
        &24.0(18.4)& &34.0(17.6) \\

 &1.2&  &6.2(3.9)& &7.6(6.2)& &8.2(12.8)&
        &9.7(19.5)& &27.0(29.9) \\

 &1.5&  &6.0(3.7)& &5.9(3.7)& &5.8(4.2)&
        &5.8(4.2)& &6.6(6.6) \\

 &2.0&  &5.7(3.1)& &5.4(3.1)& &5.3(2.9)&
        &5.4(3.0)& &5.9(3.0) \\

 &2.5&  &5.5(3.0)& &5.3(3.1)& &5.2(2.8)&
        &5.2(2.8)& &5.8(2.8) \\

 &3.0&  &5.5(2.9)& &5.5(3.1)& &5.2(2.8)&
        &5.2(2.8)& &5.8(2.8) \\

 &3.5&  &5.5(2.9)& &5.5(3.1)& &5.2(2.9)&
        &5.2(2.7)& &5.8(2.7) \\

 &7.0&  &5.8(3.1)& &6.1(3.6)& &5.2(2.9)&
        &5.2(2.8)& &5.8(2.7) \\

 &$\ln(n)/2$&  &5.6(3.0)& &5.4(3.1)& &5.2(2.8)&
               &5.2(2.8)& &5.8(2.8) \\

 &$\ln(n)$&   &5.7(3.1)& &5.8(3.3)& &52(2.9)&
               &5.2(2.8)& &5.8(2.7) \\

 &$\ln(k)/2$& &5.8(3.4)& &5.3(3.1)& &5.2(2.8)&
              &5.2(2.8)& &5.8(2.8) \\

 &$\ln(k)$&  &5.5(2.9)& &5.6(3.2)& &5.2(2.9)&
             &5.2(2.8)& &5.8(2.7) \\

\hline
\end{tabular}
\end{table}

\begin{table}[h] \caption{ Median of MSE for different initial knots with MGCV Method and the Difference of
the First and  Third Quartiles of MSE (in Parentheses) for Example
2 } \tabcolsep 0.05cm  \center
\begin{tabular}{cccc cccccccccccccccc}
\hline \hline

  & & & \multicolumn{9}{c}{Initial Knots Number} &  \\ \cline{3-12}
   &Inflation factor $\gamma$ & &{30} & & {60} & &
  {90} & & {120}  & &
  {150}\\
\hline
 &1.0& &1.08(6.7)& &12.4(8.3)& &14.6(14.3)& &16.3(21.2)&
       &2.64(32.1) \\

 &1.2& &9.9(6.2)& &10.4(6.8)& &10.3(8.3)& &10.8(8.9)&
       &10.9(9.2) \\

 &1.5& &9.3(5.8)& &9.3(5.5)& &9.1(5.5)& &9.3(5.5)&
       &9.4(5.2) \\

 &2.0& &9.0(5.3)& &9.1(5.0)& &9.0(5.1)& &9.2(5.2)&
       &9.4(5.1)  \\

 &2.5& &8.9(5.2)& &9.3(5.3)& &9.4(5.1)& &9.5(5.1)&
       &9.8(5.2) \\

 &3.0& &9.1(5.5)& &9.7(5.6)& &9.6(5.5)& &10.1(5.6)&
       &9.9(5.2) \\

 &3.5& &9.2(5.5)& &10.0(6.0)& &10.2(6.2)& &10.7(5.9)&
       &10.1(5.2) \\

 &7.0& &12.9(8.7)& &14.5(8.8)& &15.2(8.5)& &13.7(6.1)&
       &10.8(6.0) \\

 &$\ln(n)/2$&  &9.0(5.5)& &9.5(5.2)& &9.5(5.3)& &9.9(5.3)&
               &9.8(5.2) \\

 &$\ln(n)$&    &10.8(6.5)& &12.8(7.6)& &12.6(7.3)& &13.4(6.4)&
               &10.5(5.6) \\

 &$\ln(k)/2$&  &9.2(5.6)& &9.1(5.0)& &9.2(5.1)& &94(5.1)&
               &9.8(5.2) \\

 &$\ln(k)$&    &9.2(5.5)& &10.7(6.2)& &11.2(6.8)& &12.4(6.3)&
               &10.3(5.4) \\

\hline
\end{tabular}
\end{table}

\begin{table}[h] \caption{ Median$\times 1000$ of MSE for different initial knots with PREC Method and the Difference of
the First and  Third Quartiles of MSE (in Parentheses) for Example
2 } \tabcolsep 0.05cm  \center
\begin{tabular}{cccc cccccccccccccccc}
\hline \hline

  & & & \multicolumn{9}{c}{Initial Knots Number} &  \\ \cline{3-12}
   &Inflation factor $\gamma$ & &{30} & & {60} & &
  {90} & & {120}  & &
  {150}\\
\hline
 &1.0& &10.9(7.5)& &12.6(8.8)& &16.5(14.4)&
       &21.7(20.1)& &35.2(26.9) \\

 &1.2& &10.2(6.5)& &11.3(7.4)& &12.0(13.0)&
       &13.0(18.8)& &16.8(33.5) \\

 &1.5& &9.5(6.0)& &9.9(6.1)& &9.8(6.7)&
       &10.0(6.6)& &10.5(8.9) \\

 &2.0& &9.2(5.5)& &9.1(5.2)& &9.0(5.4)&
       &9.3(5.4)& &9.4(5.2) \\

 &2.5& &8.9(5.4)& &9.2(5.2)& &9.1(5.1)&
       &9.2(5.2)& &9.5(5.2) \\

 &3.0& &8.9(5.3)& &9.3(5.3)& &9.3(5.2)&
       &9.4(5.1)& &9.8(5.2) \\

 &3.5& &9.0(5.5)& &9.5(5.4)& &9.5(5.2)&
       &9.7(5.3)& &9.8(5.2) \\

 &7.0& &10.1(5.7)& &11.0(6.8)& &11.1(6.6)&
       &11.7(6.3)& &10.3(5.4) \\

 &$\ln(n)/2$& &8.9(5.3)& &9.3(5.2)& &9.2(5.1)&
              &9.3(5.1)& &9.6(5.2) \\

 &$\ln(n)$& &9.5(5.5)& &10.4(6.3)& &10.4(6.1)&
            &11.0(6.2)& &10.1(5.3) \\

 &$\ln(k)/2$& &9.3(5.7)& &9.2(5.2)& &9.0(5.1)&
              &9.2(5.3)& &9.5(5.1) \\

 &$\ln(k)$&   &9.0(5.5)& &9.7(5.7)& &10.1(6.1)&
              &10.7(5.8)& &10.1(5.2) \\
\hline
\end{tabular}
\end{table}

\clearpage

\section{Discussion and Extension}
The non-convex penalized regression spline can be easily extended
to multivariate regression models. In Chapter 2, we have applied
our non-convex penalized regression spline to a partial linear
model to reduce the modeling bias. In section, we briefly outline
how to extend our approach to nonparametric additive model.

First, we suppose that we have $J$ predictor variables and that
$\x_i=(x_{i,1},\ldots,x_{i,J})^T$ is the vector of predictor
variables for the $i$th case. The additive model considered is
defined as follows:
\begin{equation}
\label{add01} y_i=f(\x_i)+\eps_i =\mu+\Sigma_{j=1}^J
f_j(x_{i,j})+\eps_i.
\end{equation}
where $\E f_j(x_{i,j})=0$, $1 \le j \le J$, is imposed for
identifiability.  As in univariate setting, we may use spline
function $\hat{f}_j(x_{i,j},\B_j)$ of order $p$ with $K_j$ knots,
$k_{1,j},\ldots, k_{K_j,j}$, to approximate the function
$f_j(x_{i,j})$ subject to the constrains that $\sum_1^n
\hat{f}_j(x_{i,j},\B_j)=0$ where $\hat{f}_j(x_{i,j},\B_j)$ is
defined as follows,
\begin{equation}
\label{add02} \hat{f}_j(x_{i,j},\B_j)=\beta_{0,j}+
\beta_{1,j}x_{i,j}+\cdots +\beta_{p-1,j}x_{i,j}^{p-1}
+\sum\limits_{l=1}^{K_j} \beta_{p+l-1,j}(x_{i,j}-k_l)_+^{p-1}.
\end{equation}

\noindent Therefore, we obtain an additive spline model
\begin{equation}
\label{add03} \hat{f}(\x_i,\B)=\mu+\sum\limits_{j=1}^J
\hat{f}_j(x_{i,j},\B_j),
\end{equation}
to approximate the additive model (\ref{add01}), where
$$\B=(\mu, \beta_{0,1},\beta_{1,1},\ldots,\beta_{p+K_1-1,1},\ldots,\beta_{p+K_J-1,J})^T$$

The non-convex penalized criterion is to minimize
\begin{equation}
\label{add04}  \sum\limits_{i=1}^n \{y_i-\hat{f}(\x_i,\B)\}^2 +
n\sum\limits_{j=1}^J\sum\limits_{k=1}^{K_j} p_{\lambda_{n,j}}
(|w_{p+k-1,j}\beta_{p+k-1,j}|),
\end{equation}
 subject to
the constraints that $\sum_1^n \hat{f}_j(x_{i,j},\B_j)=0$, where
$p_{\lambda_{n,j}}(\cdot)$ is the non-concave penalty with the
penalty parameter $\lambda_{n,j}$. In practice, we replace
$\mu+\beta_{0,1}+\beta_{0,2}+\cdots+\beta_{0,J}$ in ( \ref{add03})
by a single parameter $\beta_0$ to release the constraints imposed
on the non-concave penalty least-squares (\ref{add04}). As the
univariate setting, we choose the SCAD as the non-concave penalty.
Now, we have transferred the additive model to a non-concave
penalty least-squares problem. Hence it is no difference between
the minimization of (\ref{add04}) and that of (\ref{s24}). The
iterative algorithm used in the univariate setting can be still
used in additive model settings.

To simplify our procedure to fit an additive spline model, we set
all penalty parameters equalling to a global penalty parameter.
Then we select this globe penalty parameter by the MGCV and the
PREC as in the univariate setting. To make our estimate for the
additive model more accurately, we can also select different value
for the different penalty parameters used in (\ref{add04}) by the
MGCV and the PREC through some optimization algorithms in a $J$
dimension space.

\end{document}